\documentclass[]{article}

\pdfoutput=1 

\usepackage{amssymb}
\usepackage{amsmath,amsfonts}
\usepackage{algorithm2e}
\usepackage{caption}
\usepackage{subcaption}
\usepackage{url}
\usepackage{cleveref}
\usepackage{eurosym}
\usepackage{todonotes}
\usepackage{xcolor}
\usepackage{pgfplots}
\usepgfplotslibrary{statistics}
\pgfplotsset{compat = newest}
\usepackage{comment}
\usepackage{tikz}
\usepackage{authblk}
\usepackage[backend=bibtex]{biblatex} 
\title{Grid-Aware Real-Time Control and Balancing Between Microgrids}
\author[1]{Jens H\"onen}
\author[1]{Johann L. Hurink}
\author[2,3]{Bert Zwart}
\affil[1]{Faculty of EEMCS, University of Twente, Enschede, The Netherlands}
\affil[2]{Department of Mathematics and Computer Science, Eindhoven University of Technology, Eindhoven, The Netherlands}
\affil[3]{Centrum Wiskunde \& Informatica (CWI), Amsterdam, The Netherlands}
\date{}

\begin{document}
	
	\maketitle
	
	\begin{abstract}
		Due to the energy transition, lots of research has been conducted within the last decade on the topics of energy management systems or local energy trading approaches, often on the day-ahead or intraday level. A large majority of these approaches focuses on 15 or 60-minute time intervals for their operation, however, the question of how the planned solutions are realized within these time intervals is often left unanswered. Within this work, we aim to close this gap and propose a real-time balancing and control approach for a set of microgrids, which implements the day-ahead solutions. The approach is based on a three-step framework, in which the first step consists of ensuring the feasibility of devices within the microgrids. The second step focuses on the grid constraints of the connecting medium voltage grid using the DC power flow formulation due to the running time requirements of a real-time approach. The last step is to propagate the solution into the individual microgrids, where the allocated power needs to be distributed among the devices and households. Within a case study, we show that the proposed real-time control approach works as intended and is comparable to an optimal offline algorithm under some mild assumptions.
	\end{abstract}
	
	\section{Introduction}\label{}
	In the last decades, the energy transition has gotten more and more attention, and the integration of renewable energy sources, such as wind power or photovoltaic (PV) systems, has increased drastically (see e.g., \cite{PVcapacityEU2021Kougiasetal}). Along with this increase of renewable energy, mainly in the form of electricity, various aspects of everyday life, such as heating or mobility, are now being electrified. This change poses a huge challenge to current electricity systems, and within the last decade, a lot of research has been carried out to tackle these challenges. \\
	
	One promising approach getting much attention is the concept of a microgrid, consisting of a set of households connected to the same part of the electricity grid (e.g., a neighborhood) (\cite{MGConcept2004Lasseteretal}). Within a microgrid, a joint (but possibly decentralized) energy management system (EMS) controls various household devices and loads to achieve some microgrid objective, which can be to minimize the total cost of the microgrid or to maximize the usage of locally produced (PV) electricity (see \cite{SurveyMGEMS2020Elmouatamidetal}, \cite{SurveyEnergyManagement2019Garciaetal}, \cite{SurveyMicrogridEMS2018Ziaetal} for recent surveys). Many of the microgrid approaches consider the operation on a day-ahead and intraday level, whereby, the considered time horizon is split up into equidistant time slots, often of length between 15 minutes and 1 hour. These approaches view the problem from an energy perspective, thereby paying no attention to the power distribution within a time slot. While this is a reasonable assumption for the (long-term) day-ahead operation and corresponding markets, in the real-time operation of the physical grid, demand, and generation are not equally spread within each time slot, and large fluctuations may appear. These fluctuations may lead to short-term mismatches between the planned energy or power exchange and the actual loads and generation, which may cause power quality problems (\cite{PowerQualityOverview2017Bollenetal}, \cite{PowerQualityBattery2018Dasetal}). Therefore, a more detailed view on the power profiles within the individual time slots is needed. This short to real-time perspective asks to consider the power aspects of the problem to be able to integrate grid constraints.\\
	
	In comparison to the mostly optimization-based EMS approaches for the day-ahead and intraday operation, real-time approaches are often based on (simple) rule-based control mechanisms to cope with the computational restrictions of real-time control. Furthermore, one of the main reasons for power quality issues in microgrids is a surplus of PV generation during times of low demand. There are two core research directions to tackle such issues. The first is to simply curtail the PV generation, while the second uses batteries to store some of the generation and thereby reduce the grid feed-in of electricity. Common approaches regarding the curtailment of (PV) generation are droop-control (\cite{VoltContDroop2011Tonkoskietal}, \cite{VoltContDroop2017Khaledianetal}, \cite{VoltContDroop2017Kallamadietal}, \cite{RTOnlineFeedbackOptACMG2023Olivesetal}) or the volt-watt scheme (\cite{VoltContvoltwatt2020Emmanueletal}). An important and challenging aspect of PV curtailment within microgrids is the fairness of the curtailment among different households (see e.g., \cite{VoltContFairness2016Latifetal}, \cite{VoltContFairness2021Vadavathietal}). When using batteries to reduce the feed-in, it is not sufficient to focus only on the current situation and corresponding actions, as capacity and possible future problems and decisions already need to be considered. To still be able to compute solutions efficiently, Lyapunov optimization is often used as an online control approach for batteries to reduce peaks in demand or PV feed-in (\cite{OnlineBattControl2023Shietal}, \cite{RTEMSinMG2017Shietal}, \cite{RTBattLoadLyapunov2018LiDong}, \cite{RTControlLyapunov2020Ahmadetal}). Other approaches use model predictive control or reinforcement learning to incorporate additional aspects, such as ancillary services or voltage constraints into real-time control schemes (\cite{RTRHAncillaryService2020Nelseonetal}, \cite{RTMPCSecondVoltage2023Escobar}, \cite{RTsafeRLEnergyHub2022Qiuetal}).  Furthermore, in \cite{VoltContOLTCBatt2021Tewarietal}, a mixture between load curtailment and battery usage is proposed and in \cite{RTp2pEM2021Guoetal} a real-time peer-to-peer energy market is considered, which is solved using a novel OC-ADMM method.\\
	
	Another interesting way to deal with real-time control of batteries and loads without having to consider the uncertainty of future time slots is to combine (long-term) energy management approaches with real-time control. This idea has only recently seen more attention (see e.g., \cite{DARTmarket2013Marzbandetal}, \cite{RTCPVstorage2018Conteetal}, \cite{DARTMIP2019Conteetal}, \cite{RTStochOptEMSRH2019Hafizetal}, \cite{RTStochOptEMS2020Hafiz}, \cite{OMEMS2021Bintoudietal}, \cite{MTPVbatt2023Lietal}). The main principle among these approaches is to use the day-ahead energy operation to guide the real-time control. To ensure feasibility, the day-ahead operation is often based on solution techniques, such as chance-constrained (\cite{RTCPVstorage2018Conteetal}) or multi-stage stochastic optimization (\cite{RTStochOptEMSRH2019Hafizetal}, \cite{RTStochOptEMS2020Hafiz}) to account for the uncertainty in PV or demand forecasts. As already mentioned, the considered time slots usually encompass 15 to 60 minutes and the most common objective is to minimize the long-term costs of the considered system. The real-time component of the approaches then tries to realize the planned solution within a single (day-ahead) time slot. Using the targets, provided by the day-ahead solution, even very simple control approaches can achieve good results. These real-time approaches can range from simple rules (\cite{DARTmarket2013Marzbandetal}, \cite{RTCPVstorage2018Conteetal}, \cite{RTStochOptEMSRH2019Hafizetal}, \cite{RTStochOptEMS2020Hafiz}) or look-up-tables (\cite{OMEMS2021Bintoudietal}) to more sophisticated mathematical optimization models (\cite{DARTMIP2019Conteetal}, \cite{MTPVbatt2023Lietal}). The objective of the real-time component usually focuses on minimizing the deviations from the planned day-ahead schedule. Another common aspect of the considered literature is that they focus on a few devices, such as a PV system, a battery, and possibly an additional (household) load. Only \cite{DARTmarket2013Marzbandetal} and \cite{MTPVbatt2023Lietal} model and control multiple (distributed) devices within the setting of a microgrid. Due to the focus on day-ahead EMSs and real-time control of only a single household or small energy system, the constraints of the underlying grid, which are usually an important aspect of real-time control, are mainly ignored. Only \cite{MTPVbatt2023Lietal} considers power flow computations in both, day-ahead and real-time. Summarizing, there is a gap in the current literature on the combination of distributed EMSs and real-time control or balancing between microgrids, whereby also grid limitations are taken into account.\\

	This work therefore aims to combine day-ahead energy operation with real-time power control to ensure feasibility of the power solution not only w.r.t. grid and device constraints, but also w.r.t. the already made decisions within the day-ahead energy layer. Hereby, we use the solution of the day-ahead stage to guide the real-time control actions. The considered setting consists of a set of connected, but independent microgrids and spans a single time slot of the day-ahead operation problem. We treat each microgrid as an active participant of the real-time control algorithm, which next to the implementation of the day-ahead plan also contributes to the feasibility of grid constraints of the connecting medium voltage (MV) distribution grid. The relatively simple (radial) structure of the low voltage (LV) grids connecting the households within a microgrid is left to the corresponding microgrids to handle. The proposed approach focuses on the real-time power control problem, whereby the main objective is to realize the planned day-ahead solution. This can be split into two aspects: The first objective is to minimize the deviations in the real-time power exchange with the market from the planned power exchange level. The second objective is to ensure that each device reaches its planned state (of charge) at the end of the time slot. To achieve these objectives, each microgrid may use the flexibility of its local devices, such as batteries or PV systems, and may also trade real-time surplus of electricity with neighboring microgrids. The key contributions of this paper are:
	\begin{itemize}
		\item We propose a general three-step framework, which allows to build and implement a real-time control approach, which uses the day-ahead energy solution as guidance for its decisions and which allows microgrids to trade and thereby support each other in achieving their planned power exchange with the market.
		\item We formulate many of the individual components of the framework as well-known combinatorial optimization problems, for which efficient algorithms exist. Thereby, we can achieve fast running times, which mainly depend on the size of the considered MV grid and the maximum number of devices per microgrid.
		\item We test and analyze the proposed real-time control approach on several cases within MV electricity grids and show that its objective value is comparable to an optimal offline solution, while still running sufficiently fast for a real-time implementation.
	\end{itemize}
	The paper is structured as follows: In Section \ref{S2Setting}, the considered setting and main motivation are presented. In Section \ref{S3Models}, the mathematical formulation of flexibility for various devices is first introduced, before presenting an aggregated microgrid and system model. In Section \ref{S4RTCA}, the three-step framework, including the different components is proposed together with possible extensions and alternatives for the individual parts. In Section \ref{S5Analysis}, the proposed real-time control approach is tested and analyzed. The work is concluded with a short summary and discussion in Section \ref{S6Conclusion}.

	\section{Setting}\label{S2Setting}
	In the following, we introduce the considered setting of both, the day-ahead as well as the real-time layer. Hereby, the main focus lies on the real-time setting.
	
	\subsection{Day-Ahead Energy Operation}
	We consider a day-ahead energy management or trading problem where the energy usage of various households and their devices within a microgrid is coordinated and planned for the next day. The corresponding approach has to ensure that the demand for each household is satisfied for that day. Hereby, device constraints, such as e.g., charging and discharging limits or capacity constraints of batteries or EVs, have to be respected. In most cases, the main objective of a corresponding EMS is to maximize the financial profit (or minimize costs) for the whole microgrid. In some cases, other or additional objectives, such as e.g., minimizing the corresponding greenhouse gas emissions or minimizing the comfort loss of households, are also considered. As the name already indicates, the day-ahead EMS runs one day before the actual realization, and the considered time horizon usually spans the whole day, although there are also cases, in which a larger time horizon is considered. For the operation, the time horizon is split up into non-overlapping time slots (in general of length 15 to 60 minutes). For each such time slot, an energy schedule has to specify how much energy each device either consumes or produces during that time slot. In addition, for each time slot, the amount of produced and bought energy has to be equal to the amount of consumed and sold energy. Hence, a solution of the day-ahead EMS consists of a set of energy profiles, which defines the planned energy usage of each device for each time slot, while always maintaining the balance between demand and supply.\\
	
	However, due to the time distance between decision-making and realization, deviations from the planned (household) demand or (PV) production occur. In addition, some (necessary) simplifications are included in the model due to the discrete time slot structure. These simplifications may cause problems in the real-time implementation of the day-ahead solution. If for instance, the forecast for the PV generation and the household demand for a certain time slot are equal, then from the day-ahead perspective, the solution to use the PV generation to satisfy the household demand is a feasible solution. When zooming into this specific time slot, however, in general, neither the household demand nor the PV generation are equally spread among the 15 minutes, and (large) differences and mismatches in their power profiles may appear. These differences may pose serious problems to the electricity system, as the system based on the day-ahead solution assumes that it has no energy exchange with the household. Taking such short-term fluctuations already in the day-ahead stage into account requires a finer time granularity, which results in two major problems. The first one concerns the complexity and size of the resulting optimization model, which may not be computationally tractable any longer. The second problem concerns the required data. It is well known that demand and production for larger time windows are much easier to predict and that the resulting values are more accurate, as short-term fluctuations within the time window may cancel out each other. On the other hand, the required time granularity has to be short enough to capture the occurring fluctuations in household demand and PV generation. However, it is hard or even impossible to predict these values accurately over a longer time horizon. Based on this, it is not feasible to alter the day-ahead EMS approach to also be able to capture the short-term fluctuations of demand and production. Therefore, we propose to add a real-time control approach, which deals with the short-term power differences in production and consumption in an online way.\\
	
	In the remainder of this work, we assume that a day-ahead solution is already given and therefore we treat the actual EMS as a black box. In addition, we assume that the EMS is robust against the uncertainty of the PV generation by using techniques such as, e.g., (adaptive) robust optimization or stochastic programming. We derive and explain this requirement in more detail in Section \ref{S5Analysis}.
	
	\subsection{Real-Time Power Control}
	
	In the real-time domain, the microgrid has to realize its planned day-ahead schedule taking into account the realizations of PV generation and the household base load. Hereby, the focus is on a single day-ahead time slot of e.g., 15 minutes. For the real-time control, this interval is split up into even smaller time slots, denoted by time slices, of length $\Delta t$. We denote the set of time slices by $\mathcal{T}$. For each of these time slices, demand needs to be equal to supply, whereby the actual realizations of PV and household load are revealed only at the beginning of their corresponding time slice. Therefore, the real-time control algorithm has to work in an online fashion, only considering (and knowing information about) the current time slice, as well as past decisions. The made decisions for the current time slice are then realized and the algorithm proceeds with the next time slice. As the goal is to realize the day-ahead solution as closely as possible, we use the planned schedule to guide the online real-time control algorithm in the right direction. In the following, we first introduce the required information from the day-ahead solution, as well as external parameters for this process:\\
	
	\begin{itemize}
		\item Planned market decisions: $\Tilde{x}^M$ denotes the planned aggregated amount of energy bought or sold at the day-ahead and intraday markets for the considered time slot.
		\item Planned Battery decisions: $\Tilde{E}^{B,k}$ denotes the planned energy within battery $k$ at the end of the time slot, and $E^{B,k}_{0}$ denotes the initial energy within battery $k$ at the start of the considered time slot. Furthermore, $\eta$ denotes the charging and discharging efficiency of the battery and $L^{B,k,C}$ ($L^{B,k,D}$) the charging (discharging) power limit. 
		\item Planned EV decisions: $\Tilde{E}^{EV,h}$ denotes the planned energy within EV $h$ at the end of the time slot, and $E^{EV,h}_{0}$ denotes the initial energy within EV $h$ at the start of the considered time slot. Furthermore, $\eta$ denotes the charging and discharging efficiency of the EV and $L^{EV,h,C}$ ($L^{EV,h,D}$) the charging (discharging) power limit.
		\item Actual household demand: $P^{HL}$ denotes the (aggregated) household load power profile of a microgrid during the considered time slot, and $P^{HL}_{t}$ denotes the expected load of time slice $t$, which is only revealed just before time slice $t$.
		\item Actual PV generation: $P^{PV}$ denotes the (aggregated) PV generation power profile of a microgrid during the considered time slot, and $P^{PV}_{t}$ denotes the realized PV generation of time slice $t$, which is only revealed just before its corresponding time slice $t$. 
	\end{itemize}
	
	In this work, we aim to precisely follow the planned schedule of the batteries and EVs on the time slot level. Hence, the microgrid has to ensure that the actual energy within the batteries and EVs at the end of the time slot is equal to the planned amount $\Tilde{E}^{B,k}$, respectively $\Tilde{E}^{EV,h}$. However, as the aggregated realized PV generation, as well as household load, may differ from the predicted amount based on which the day-ahead solution was calculated, we do not require that the actual energy exchange of the microgrid with the markets is equal to the planned energy exchange. Nevertheless, we try to minimize the sum of squared differences between the realized power exchange profile with the market and the planned power exchange level. Given the constant planned power exchange level, this results in a power exchange profile as flat as possible, which thereby also minimizes short-term fluctuations in power delivery or feed-in. \\

	To achieve the specified objectives, the microgrid can use the flexibility of its devices, in particular of the batteries and EVs, as well as the option of PV curtailment. If this flexibility is not sufficient to keep the profile of the market exchange constant throughout the whole time slot, there are two options. The first one is to trade power with neighboring microgrids, which may have flexibility left. Thereby, the goal is to keep the power exchange with the market at the planned level. If the flexibility of other microgrids is still not sufficient, the power exchange with the markets may be increased (or decreased) to ensure a balance between demand and supply at the cost of fluctuations in the power exchange with the markets.\\

	Both, the trading with other microgrids as well as the adaption of the exchange with the market, impacts the underlying electricity grid. Therefore, we have to ensure that the grid limitations are respected by the updated electricity trading. We do so by means of power flow computations, which are based on information on the underlying electricity grid. For the sake of simplicity, we assume that the connecting MV grid has a single connection to the main grid, and thereby to the electricity markets.\\

	\section{Mathematical Modeling}\label{S3Models}
	
	Based on the above-introduced setting and goal of the real-time control approach, we first focus on the modeling of a single microgrid. We present an approach to project the flexibility of devices for the whole 15-minute time slot onto the current time slice $t$ while ensuring feasibility w.r.t. the day-ahead targets of the devices. We then proceed with an aggregation of all flexibility within a microgrid and present the microgrid model for a single time slice. The second focus of this section is on a model on the system level, which merges multiple microgrids. For this, we combine the individual microgrid models and add constraints modeling the peer-to-peer trading as well as the power flow constraints.

	\subsection{Microgrid Model}
	In the following, we focus on the mathematical modeling of devices and the objective of a microgrid $i$ for a single time slice $t$. Due to the very short time length of a time slice, denoted by $\Delta t$ (e.g., 1-60 seconds), we model the variables, constraints, and the objective using power as the main unit. We thereby deviate from the energy modeling perspective of the day-ahead and intraday operation problem.

	\subsubsection{Device Flexibility}
	In this section, we present a way to model the use of flexibility of devices for a single time slice, while ensuring feasibility of the day-ahead solution. We use the battery flexibility as the main example for devices, that connect multiple time slices with each other. Furthermore, we shortly specify how the flexibility can be modeled for the remaining considered devices.\\
	
	Based on the standard multi-time slice formulation of a battery, we derive upper and lower bounds on the device flexibility for a single time slice. For the sake of simplicity, we omit the indices denoting microgrid $i$ and battery $k$. The standard multi-time slice constraints for the operation of a battery are
	\begin{align}
		0 \leq E^{B}_0 + \eta \sum_{s=1}^t x^{B,C}_{s}\Delta t - \dfrac{1}{\eta} \sum_{s=1}^t x^{B,D}_{s}\Delta t & \leq C^B \quad \forall t \in \mathcal{T},\label{mMTSB1}\\
		0 \leq \eta x^{B,C}_{t} & \leq L^{B,C} \quad \forall t \in \mathcal{T},\label{mMTSB2}\\
		0 \leq \dfrac{1}{\eta} x^{B,D}_{t} & \leq L^{B,D}\quad \forall t \in \mathcal{T},\label{mMTSB3}\\
		E^{B}_0 + \eta \sum_{s=1}^T x^{B,C}_{s}\Delta t - \dfrac{1}{\eta} \sum_{s=1}^T x^{B,D}_{s}\Delta t & = \Tilde{E}^B,\label{mMTSB4}
	\end{align}
	where $x^{B,C}_{t}$($x^{B,D}_{t}$) corresponds to the external charging (discharging) power during time slice $t$. Constraint (\ref{mMTSB1}) ensures that for all time slices, the energy within the battery is between 0 and its capacity $C^B$. Constraints (\ref{mMTSB2}) and (\ref{mMTSB3}) impose given (internal) bounds on the charging and discharging power, while constraint (\ref{mMTSB4}) ensures that the energy within the battery at the end of the time slot is equal to the planned energy $\Tilde{E}^B$.\\ 
	
	To reformulate these constraints into a single-time slice model for time slice $t$, we first reformulate constraints (\ref{mMTSB1}) and (\ref{mMTSB4}). Let $E^B_{t}$ denote the energy within the battery just before time slice $t$. Constraint (\ref{mMTSB1}) for time slice $t$ then simplifies to
	\begin{equation}\label{mSTSB1}
		0 \leq E^B_{t} + \eta x^{B,C}_{t}\Delta t - \dfrac{1}{\eta} x^{B,D}_{t}\Delta t \leq C^B .
	\end{equation}
	Constraints (\ref{mMTSB2}) and (\ref{mMTSB3}) already consider only variables for time slice $t$. The only remaining issue for the single-time slice model is to guarantee that the planned energy $\Tilde{E}^B$ is achieved at the end of the time slot. One simple approach is to backlog how much energy has to be within the battery at the end of time slice $t$ to still be able to achieve the required energy $\Tilde{E}^B$ by the end of the time slot. Hence, for the charging and discharging decision of time slice $t$, we have the following restriction on the energy within the battery at the end of the time slice:
	\begin{equation}
		\Tilde{E}^B - (T-t)L^{B,C}\Delta t \leq E^B_t + \eta x^{B,C}_{t}\Delta t - \dfrac{1}{\eta} x^{B,D}_{t}\Delta t \leq \Tilde{E}^B + (T-t)L^{B,D}\Delta t.
	\end{equation}
	We can now differentiate between charging and discharging and thereby receive the following bounds:
	\begin{equation}\label{mSTSB3}
		\dfrac{1}{\eta}(\Tilde{E}^B/\Delta t - E^B_{t}/\Delta t - (T-t)L^{B,C}) \leq x^{B,C}_{t} \leq \dfrac{1}{\eta} (\Tilde{E}^B\Delta t - E^B_{t}\Delta t + (T-t)L^{B,D}), 
	\end{equation}
	\begin{equation}\label{mSTSB4}
		-\eta(\Tilde{E}^B/\Delta t- E^B_{t}/\Delta t + (T-t)L^{B,D}) \leq x^{B,D}_{t} \leq -\eta (\Tilde{E}^B/\Delta t - E^B_{t}/\Delta t - (T-t)L^{B,C}). 
	\end{equation}
	
	In case the lower bound of either constraint (\ref{mSTSB3}) or (\ref{mSTSB4}) is strictly positive, we are forced to charge, respectively discharge to still be able to achieve the planned energy at the end of the time slot. In case both lower bounds are negative, together with constraints (\ref{mMTSB2}), (\ref{mMTSB3}) and (\ref{mSTSB1}) we simply can use a single battery variable $x^B_t$ with the following bounds:
	\begin{align}
		l^B_t &= \eta \max \{ - E^B_t/\Delta t, -L^{B,D}, ((\Tilde{E}^B - E^B_{t})/\Delta t - (T-t)L^{B,C})\},\label{lflexbatt}\\
		u^B_t &= \dfrac{1}{\eta} \min \{ (C^B-E^B_t)/\Delta t, L^{B,C}, ((\Tilde{E}^B_t - E^B_{t})/\Delta t + (T-t)L^{B,D}) \}.\label{uflexbatt}
	\end{align}
	
	The EV flexibility can be treated the same way as the battery flexibility since the considered time horizon is just a single day-ahead time slot for the real-time approach, and therefore, each EV is considered to be either available and can be charged and discharged, or is not available, which renders the flexibility to 0.\\
	
	The PV generation and the load-supply balance constraints can be modeled in a straightforward manner, as these only contain variables of the current time slice $t$, leading to the following device flexibility constraint for time slice $t$:
	\begin{equation}
		0 \leq x^{PV}_{t} \leq P^{PV}_{t},
	\end{equation}
	for the PV generation, and 
	\begin{equation}
		x^M_{t} + x^{PV}_t -\sum_{k \in \mathcal{N}_{B}}x^B_{k,t} - \sum_{h \in \mathcal{N}_{EV}}x^{EV}_{h,t} = P^{HL}_{t},
	\end{equation}
	for the supply-demand-balance constraint of the microgrid for time slice $t$. Hereby, $N_B$ ($N_{EV}$) represents the set of batteries (EVs) within the considered microgrid.

	\subsubsection{Aggregation of Flexibility}\label{S312Aggregation}
	The flexibility model of a microgrid for time slice $t$ can be further simplified by combining the flexibility of all devices within the microgrid into a single variable $x^D_t$: 
	\begin{equation}
		x^D_t = P^{HL}_t - x^{PV}_t + \sum_{k \in \mathcal{N}_{B}}x^{B,k}_t + \sum_{h \in \mathcal{N}_{EV}}x^{EV,h}_t.
	\end{equation}
	The corresponding upper and lower bounds of the resulting single device variable are then 
	\begin{equation}
		u^D_{t} = P^{HL}_{t} + \sum_{k \in \mathcal{N}_{B}}u^{B,k}_{t} + \sum_{h \in \mathcal{N}_{EV}}u^{EV,h}_{t},
	\end{equation}
	and 
	\begin{equation}
		l^D_{t} = P^{HL}_{t} - P^{PV}_{t} + \sum_{k \in \mathcal{N}_{B}}l^{B,k}_{t} + \sum_{h \in \mathcal{N}_{EV}}l^{EV,h}_{t},
	\end{equation}
	where $u^{B,k}_{t}$ and $u^{EV,h}_{t}$, respectively $l^{B,k}_{t}$ and $l^{EV,h}_{t}$, correspond to the upper (lower) limits of the device flexibility of battery $k$ and EV $h$ as derived in (\ref{lflexbatt}) and (\ref{uflexbatt}). Note, that the (aggregated) household load of a microgrid, $P^{HL}_t$, can be seen as a variable with lower and upper bounds of $P^{HL}_t$. This aggregation of individual device flexibility leads to the following constraint
	\begin{equation}
		l_t^D \leq x^D_t \leq u^D_t.
	\end{equation}
	
	The above aggregation is also an advantage regarding the privacy of data and information of individual devices and households within the microgrid. Due to the aggregation, no detailed, individual device information can be accessed by involved parties outside the microgrid.

	\subsubsection{Single-Time Slice Microgrid Model}
	As explained in Section \ref{S2Setting}, the objective of the microgrid is to achieve a flat power exchange profile with the market. Hence we introduce the market exchange variable $x^M_{i,t}$ for microgrid $i$ and time slice $t$. Due to the focus on a single time slice at each iteration, we introduce a parameter $X_i$ for each microgrid, which represents the desired power exchange level with the markets. To achieve a flat profile, we initially set the value to the average power required throughout the whole time slot to satisfy the planned energy exchange with the market. To reach this value $X_i$, microgrid $i$ may use its internal (device) flexibility, as derived in Section \ref{S312Aggregation}. However, the flexibility may not be sufficient to achieve the desired exchange level. In such a case, microgrid $i$ may also directly trade with neighboring microgrids $j$ and thereby use the peer-to-peer trading component. Let $\mathcal{MG}$ denote the set of all microgrids, and let $x_{i,j,t}^{P2P}$ denote the power traded from microgrid $i$ to microgrid $j$ ($i,j \in \mathcal{MG}$), whereby positive values represent an import of power from $j$ to $i$ and negative values an export. Combining all three aspects, we have the following single-time slice microgrid model for microgrid $i$ and time slice $t$:
	\begin{align}
		\min & (x^M_{i,t} - X_i)^2\\
		\text{s.t. } &l^{D}_{i,t} \leq x^D_{i,t} \leq u^D_{i,t},\\
		& x^M_{i,t} + \sum_{j \in \mathcal{MG}} x_{i,j,t}^{P2P} = x^D_{i,t}.
	\end{align}
	
	\subsection{System Model}
	Within the overall system model, we consider a set of individual microgrids that are connected by an electricity grid structure. To model this overall system, we need to add further constraints to align the peer-to-peer decisions between microgrids and to model the power flow computations for the grid connecting the microgrids. 
	
	\subsubsection{Peer-to-Peer Trading}
	Within the single microgrid model, the peer-to-peer trading variables $x_{i,j,t}^{P2P}$ have already been introduced and used within the demand-supply balance constraint. However, in the system model, connecting the models of various microgrids with each other, we need to ensure that the power microgrid $i$ sends to microgrid $j$ is consistent with what microgrid $j$ receives from microgrid $i$. Therefore, we introduce the following constraint
	\begin{equation}
		x_{i,j,t}^{P2P} + x_{j,i,t}^{P2P} = 0 \quad \forall i,j \in \mathcal{MG}.
	\end{equation}
	
	\subsubsection{Power Flow}
	Let the underlying MV grid connecting the given microgrids consist of a set of buses, denoted by $\mathcal{N}$, and a set of lines $\mathcal{L} \subset \mathcal{N}\times\mathcal{N}$, connecting the buses. We assume, that each microgrid can be linked to a bus in the electricity grid, implying that $\mathcal{MG} \subseteq \mathcal{N}$. Let $L^{\max}_{i,j}$ denote the thermal limit of the line $(i,j) \in \mathcal{L}$ and let $x_{i,j}$ denote its reactance. Both, $L^{\max}_{i,j}$ and $x_{i,j}$ are parameters, specifying characteristics of the lines of the electricity grid and are assumed to be known.\\
	
	Based on decades of research on power flow computations, various power flow formulations have been developed and analyzed in detail (see e.g., \cite{OPFconicRelax2020Zohrizadehetal}, \cite{OPFhistory2012Maryetal}). Due to the strict time requirements of a real-time control approach, as well as the possibility to derive an analytical solution approach, we focus on the DC power flow formulation. In addition, this formulation can be applied to a wide range of grid topologies, which fits well with the considered MV grid, which can range from radial to ring or meshed structures. The DC power flow formulation is an approximation of the AC power flow equations, \cite{DCACOPF2021Baker}, and is derived based on three main assumptions and simplifications:
	\begin{enumerate}
		\item The resistance of the lines is negligible.
		\item The bus voltage magnitudes are approximately 1.
		\item The voltage angle difference $\delta_{i,j}$ for lines $(i,j)$ are small and thereby $\cos{(\delta_{i,j})}\approx 1$ and $\sin{(\delta_{i,j})}\approx\delta_{i,j}$.
	\end{enumerate}
	These assumptions simplify the original (AC) power flow equations significantly by removing some variables and constraints, leading to a set of linear constraints, which can be solved efficiently. However, it should be noted that due to the simplification, only the thermal capacities of the network are considered and voltage constraints are ignored.\\
	
	Within this system model, we have to ensure that the power flows within the network, resulting from the trading decisions of the microgrids, respect the given network constraints. For this, the resulting power flows within the grid have to be determined.\\
	
	$p^L_{i,j}$ denotes the real power flow in the line $(i,j)\in \mathcal{L}$ and let $\theta_i$ be the voltage angle at bus $i$. These are the two variables in the model and they are strongly connected to the power generation and consumption in the buses of the grid. The basic DC power flow equations are:
	\begin{equation}\label{eqDC1}
		p^L_{i,j} = \dfrac{1}{x_{i,j}}(\theta_i - \theta_j) \quad \forall (i,j) \in \mathcal{L}.
	\end{equation}
	Next to these power flow laws, flow balance constraints link the power flow over lines to the generation and consumption at the buses:
	\begin{equation}\label{eqDC2}
		x^D_{i} = \sum_{(i,j) \in \mathcal{L}} p^L_{i,j} \quad \forall i \in \mathcal{N}.
	\end{equation}
	W.l.o.g., we define $x^D_i=0$ for all buses, which are neither a microgrid nor the market (i.e. for all $i \in \mathcal{N} \setminus \mathcal{MG}$). To limit the power flow over a line $(i,j)$ to its thermal limits, we add the following constraint:
	\begin{equation}
		-L^{\max}_{i,j} \leq p^L_{i,j} \leq L^{\max}_{i,j} \quad \forall (i,j) \in \mathcal{L}.
	\end{equation}
	
	\subsubsection{Single-Time Slice System Model}
	Based on the presented microgrid model as well as the power flow constraints and the peer-to-peer constraints, the overall system model for a single time slice $t$ is given by: 
	
	\begin{align}
		\min & \sum_{i \in \mathcal{MG}} (x^M_{i,t}-X_i)^2\label{stssmObj} \\
		\textit{s.t. } & l^D_{i,t} \leq x^D_{i,t} \leq u^D_{i,t} \quad \forall i \in \mathcal{MG},\\
		& x^M_{i,t} + \sum_{j \in \mathcal{MG}} x_{i,j,t}^{P2P} = x^{D}_{i,t} \quad \forall i \in \mathcal{MG},\\
		& x_{i,j,t}^{P2P} + x_{j,i,t}^{P2P} = 0 \quad \forall i,j \in \mathcal{MG},\label{smP2P}\\
		& p^L_{i,j,t} = \dfrac{1}{x_{i,j}}(\theta_i - \theta_j) \quad \forall (i,j) \in \mathcal{L},\\
		& x^D_{i,t} = \sum_{(i,j) \in \mathcal{L}} p^L_{i,j,t} \quad  \forall i \in \mathcal{N},\\
		& -L^{\max}_{i,j} \leq p^L_{i,j,t} \leq L^{\max}_{i,j} \quad \forall (i,j) \in \mathcal{L}.\label{stssmLastConstr}
	\end{align}
	In the next section, we present an efficient algorithm to solve this optimization problem.

	\section{Real-Time Control Algorithm}\label{S4RTCA}
	In the following, we first give a high-level view of the algorithm, before explaining the individual steps in detail afterwards. After the explanation, we provide an outlook on possible extensions or alternatives to solve this problem.
	
	\subsection{Three-Step Framework}\label{S4.1Alg}
	The high-level idea of the real-time control algorithm can be split up into three main steps: 
	\begin{enumerate}
		\item The first main step is to create a device-feasible solution. We first assign to each microgrid its 'ideal' power exchange with the market, hereby not taking into account whether this solution is feasible w.r.t. the device flexibility or not. Afterward, each microgrid communicates either its upward and downward flexibility or its surplus or demand in case the ideal power exchange cannot be realized with its flexibility. If at least one microgrid is device infeasible, the peer-to-peer trading option between microgrids and the option to increase or decrease the power exchange with the market is enabled. To determine corresponding trades between microgrids, a min-cost flow problem is formulated, which can be solved efficiently.
		\item Based on the resulting device-feasible solution, the second step is to ensure that the solution is also grid feasible. Using power flow equations, we compute the power flow on the lines of the electricity grid. If any thermal line limit is exceeded, a repair algorithm enables additional trades between microgrids and the market to achieve a grid-feasible solution, which maintains the feasibility w.r.t. the device flexibility. The repair algorithm is based on power transmission distribution factors (PTDFs), which are a reformulation of the classical DC power flow equations (\ref{eqDC1}) and (\ref{eqDC2}). The problem is then formulated as a simple quadratic problem (QP) with linear constraints, which can also be solved efficiently.
		\item The resulting solution is both, grid and device feasible, and is then communicated to each microgrid in the third step. Each microgrid can then choose its own way to allocate its corresponding power among the available devices.
	\end{enumerate}
	
	Algorithm \ref{algRTC} depicts the real-time control algorithm over a given time slot, split into a set $\mathcal{T}$ of time slices. The control algorithm is based on the three-step framework, which is deployed for every time slice $t \in \mathcal{T}$.
	
	\begin{algorithm}
		\caption{Three-Step Real-Time Control}\label{algRTC}
		\For{$t \in \mathcal{T}$}{
			1. create initial solution $x^M_{i,t}=X_i$ for each microgrid $i$\;
			\If{$x^M_{i,t} \notin [l^D_{i,t}, u^D_{i,t}]$ for some $i \in \mathcal{MG}$}{
				Enable P2P trading with other microgrids and the markets by solving \textbf{Min-Cost Flow Problem}\;
			}
			2. Calculate power flow\;
			\If{Line limits are exceeded}{
				Repair solution by solving \textbf{Repair Problem}\;
			}
			3. Update parameters\;
		}
	\end{algorithm}
	
	In the following, we explain each step in detail.
	
	\subsection{Device Feasibility}\label{S4.2DF}
	The main goal of the first step is to create a solution for the current time slice $t$, which does not violate any device constraints. The secondary objective is to keep the power exchange with the market as close as possible to the given value $X_i$ for each microgrid. Therefore, the algorithm creates a first solution by assigning each microgrid its ideal power exchange $X_i$. If this power exchange is feasible for each microgrid, the algorithm continues with the second step, namely checking the grid constraints. However, if for some microgrid $i$, $x^M_{i,t} = X_i \notin [l^D_{i,t}, u^D_{i,t}]$, then we enable the peer-to-peer trading between microgrids and the market to reach a device feasible solution.
	
	\subsubsection{Min-Cost Flow Problem}
	Given the initial solution for each microgrid, we solve a min-cost flow problem to reach device feasibility. The resulting flow can then be translated into peer-to-peer trades between microgrids (and the market). We first split up the set of microgrids into three sets, which form the basis of the graph, on which the min-cost flow problem is solved.\\
	
	Let $\mathcal{D}= \{ i \in \mathcal{MG}: X_i \leq l^D_{i,t} \} $ be the set of microgrids, for which the ideal power is not sufficient to satisfy their minimal demand, let $\mathcal{S} = \{ i \in \mathcal{MG}: X_i \geq u^D_{i,t} \}$ be the set of microgrids, where the maximal device flexibility is not sufficient to consume the ideal power, and let $\mathcal{F} = \mathcal{MG} \setminus \{\mathcal{D} \cup \mathcal{S} \}$ denote the set of remaining microgrids, which are device feasible. Additionally, define $x(\mathcal{D}) = \sum_{i \in \mathcal{D}} l^D_i - X_i$ to be the total demand, which still needs to be provided and $x(\mathcal{S}) = \sum_{i \in \mathcal{S}} X_i - u^D_i$ to be the total surplus from microgrids in $\mathcal{S}$.\\
	
	\begin{figure}
		\centering
		\includegraphics[width=0.50\textwidth]{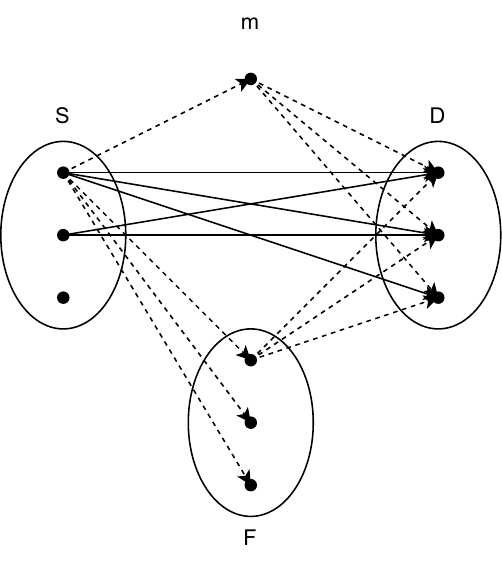}
		\caption{Sketch of the graph consisting of the microgrid sets $\mathcal{S}$, $\mathcal{D}$, $\mathcal{F}$ and $m$. The solid arcs represent connections which are always part of the graph, while the dotted ones depend on the case.}
		\label{fig:MCF}
	\end{figure}
	
	The main idea is to use the microgrids in $\mathcal{S}$ as sources of flow, the microgrid in $\mathcal{D}$ as sinks, and microgrids in $\mathcal{F}$ and the market as both, sources or sinks, depending on the current needs. There are three different situations regarding the total demand or surplus of the whole set of microgrids. If $x(\mathcal{S}) - x(\mathcal{D}) >0$, we have a surplus of power and need to distribute the remaining part among the microgrids in $\mathcal{F}$ or the market, which then act as sinks. If $x(\mathcal{S}) - x(\mathcal{D})<0$, then the microgrids in $\mathcal{S} \cup \mathcal{D}$ are not able to satisfy their total consumption and we need to distribute power from microgrids in $\mathcal{F}$ to microgrids in $\mathcal{D}$ or buy additional power from the market. If $x(\mathcal{S}) - x(\mathcal{D}) =0$, we can restrict the peer-to-peer trading scheme to microgrids in $\mathcal{D}$ and $\mathcal{S}$.\\
	
	Depending on the concrete case, the graph is slightly modified to allow classical min-cost flow algorithms to solve the problem efficiently. These modifications contain the introduction of a single main source and sink to model the flexibility of microgrids in $\mathcal{F}$ and the market, as well as costs on the arcs connecting the various microgrids with each other. These arc costs are defined in such a way, that trades between microgrids in $\mathcal{S}$ and $\mathcal{D}$ are preferred over any other trade. Furthermore, trades with microgrids in $\mathcal{F}$ are still preferred over trades with the market $m$ to ensure that in the resulting solution, the flexibility within the microgrids is first used up before increasing or decreasing the power exchange with the market.\\
	
	The resulting solution of the min-cost flow problem specifies the trading between the various microgrids. In detail, the flow over an arc $(i,j)$ can be seen as the power traded from microgrid $i$ to microgrid $j$. Based on this solution, we can compute the power consumed or produced by each microgrid for time slice $t$, which then can directly be used as input for the power flow computation.\\
	
	\subsection{Grid Feasibility}\label{S4.3GF}
	Based on the solution of the previous step, we check whether the planned solution is also feasible w.r.t. the grid constraints. The algorithm is based on well-established power flow equations to compute the power flow on the lines of the grids. If these turn out to be infeasible, a repair algorithm adjusts the planned solution to reach grid feasibility.

	\subsubsection{Power Flow Computation}
	Due to the computational requirement, we make use of DC power flow equations (as already introduced in Section \ref{S3Models}) to calculate the resulting power flow $p^L_l$ for each line $l \in \mathcal{L}$ of the considered electricity grid. If for no line the corresponding thermal limit $L^{max}_l$ is exceeded, we can implement the microgrid solution and proceed with step three (see Section \ref{S4.4PU}). If some line limits are violated, we need to adjust the device-feasible solution using a repair algorithm to achieve a grid-feasible solution.

	\subsubsection{Repair Algorithm}
	The main variables for the repair problem are additional trades between microgrids ($i$ and $j$), denoted by $\Delta x_{i,j}$. In a first step, we need to determine the remaining device flexibility of each microgrid, which can be used for the additional trades. For microgrid $i$ and time slice $t$,
	\begin{equation*}
		l'_i = l^D_{i,t} - \hat{x}^D_{i,t},
	\end{equation*}
	is the lower limit of flexibility, and
	\begin{equation*}
		u'_i = u^D_{i,t} - \hat{x}^D_{i,t},
	\end{equation*}
	is the upper limit of flexibility for additional trades, whereby $\hat{x}^D_{i,t}$ corresponds to the solution of the previous step. This leads to the following constraint for the repair problem: 
	\begin{equation}\label{mRflex}
		l'_i \leq \sum_{j \in \mathcal{MG}} \Delta x_{i,j} \leq u'_i \quad \forall i \in \mathcal{MG}\cup \{ m \}.
	\end{equation}
	Hereby, index $m$ denotes the market, for which we assume infinite upper and lower bounds. To ensure that the power traded from microgrid $i$ to microgrid $j$ equals the negative of what $j$ receives from $i$, we add the constraint
	\begin{equation}\label{mRtradeeq}
		\Delta x_{i,j} + \Delta x_{j,i} = 0 \quad \forall i,j \in \mathcal{MG}\cup \{ m \}.
	\end{equation}
	Depending on the physical properties and the structure of the underlying electricity grid, the power flow between two microgrids does not necessarily take a single path but may spread among various paths, connecting the two microgrids, see \cite{PTDFp2p2020Khorasanyetal}. To efficiently compute the impact of a trade on the power flow throughout the grid, we use power transmission distribution factors (PTDF), \cite{PTDFcongestion2004LiuGross}. These values approximate the change in power flow over a line given a change in power generation and consumption in certain nodes in the grid. The PTDF is closely related to the DC power flow equations and is independent of the actual power generation and demand of the microgrids. It only depends on the physical properties and the structure of the underlying electricity grid, and can thereby be computed upfront. Let $\varphi^l_{i,j}$ denote the PTDF for line $l$ and a trade between microgrids $i$ and $j$. Then the additional trades $\Delta x_{i,j}$ change the power flow on line $l$ by:
	\begin{equation}\label{mRflowcomp}
		\Delta p^L_l = 0.5 \sum_{i,j \in \mathcal{MG}} \varphi^l_{i,j} \Delta x_{i,j} \quad \forall l \in \mathcal{L}.
	\end{equation}
	Note that the factor of 0.5 stems from the sum, where we count every trade twice (once from $i$ to $j$ and once from $j$ to $i$). To ensure feasibility w.r.t. the line limits, we have to fulfill
	\begin{equation}\label{mRlinelimit}
		-L^{max}_l \leq p^L_l + \Delta p^L_l \leq L^{max}_l \quad \forall l \in \mathcal{L}.
	\end{equation}
	Constraints (\ref{mRflex}) and (\ref{mRtradeeq}) define the set of feasible trades between microgrids, constraint (\ref{mRflowcomp}) models the consequence of the additional trades, and constraint (\ref{mRlinelimit}) further restricts the trades to ones, that result in a grid-feasible solution. Note that constraints (\ref{mRflowcomp}) and (\ref{mRlinelimit}) may also be merged into one set of constraints, in which $\Delta p^L_l$ is not directly modeled.\\
	
	The main objective of the repair problem is to ensure a grid-feasible solution, which is given by constraint (\ref{mRlinelimit}). Hence, other, secondary, objectives may be added to decide which of the feasible solutions to choose. Within this work, we chose to reduce large additional trades as much as possible, which leads to the following objective:
	\begin{equation}
		\min \sum_{i,j \in \mathcal{MG}} (a_{i,j} \Delta x_{i,j})^2,
	\end{equation}
	where $a_{i,j}>0$ is an additional parameter that may indicate an ordering in preference over the additional trades, such as e.g., trades with the main grid being discouraged by a large $a$-value.\\
	
	The resulting optimization problem is a quadratic optimization problem with linear constraints. Due to the positive $a$-values in the objective, the resulting matrix is positive definite. Coupled with the linear constraints, this problem can be solved efficiently with modern solvers. The resulting solution is composed of additional trades, which together with the device-feasible solution now build up the final solution, which is feasible w.r.t. device and grid constraints.

	\subsection{Parameter Updates}\label{S4.4PU}
	The updated solution of the repair algorithm respects all grid and device limits and can now be translated into control actions for the various devices within each microgrid. The achieved market and peer-to-peer solution specifies how much power has to be distributed among the households and devices of each microgrid. In the following, we present one possible way to distribute this among the devices of microgrid $i$.\\
	
	Let $\Bar{x}^D_i$ be the power assigned to microgrid $i$ in the device and grid feasible solution for the current time slice. In a first step, each household within the microgrid receives its (inflexible) household load. We then subtract the sum of the household loads $P^{HL}_i$ from the power assigned to the microgrid to determine the remaining power to be distributed among the remaining flexible microgrid and household assets. Let $\Tilde{x}^D_i = \Bar{x}^D_i - P^{HL}_i$ denote this remaining power. There are many ways how to distribute this remaining power among the flexible devices. In our case, we want to minimize PV curtailment and distribute the resulting power equally among all batteries and EVs subject to the device constraints. Therefore the problem can be seen as a resource allocation problem, for which efficient algorithms, such as the cave-filling algorithm, \cite{CFP2016Naiduetal}, exist. The resulting solution is then used to distribute the power among the microgrid and household devices and to compute the new amount of energy in the various storage devices.\\
	
	Another parameter, which may need to be updated is $X_i$, indicating the desired power to be bought (or sold) from the market. The main goal is to keep the power profile as flat as possible for each microgrid, while also trying to stay as close as possible to the day-ahead energy solution. Hence, we start with $X_i = (\Tilde{x}^M_i/\vert \mathcal{T}\vert) \Delta t$, which corresponds to the power level of an equally spread day-ahead solution. Throughout the real-time control, deviations in market exchange from this desired level $X_i$ may appear, and the value has to be adapted. The main idea in updating $X_i$ is to equally spread the remaining amount of energy to exchange with the market among the remaining time slots. Let $X^M_{i,t-1}$ denote the total amount of energy exchanged with the market up to and including time slice $t-1$. Then, for time slice $t$, define the desired power level to be 
	\begin{equation}\label{eqPU4}
		X_i = \dfrac{\Tilde{x}^M_i-X^M_{i,t-1}}{\vert \mathcal{T}\vert-t+1}\Delta t.
	\end{equation}
	
	Based on this distribution of power among the devices and the new target level $X_i$, the flexibility of the next time slice $t+1$ can be computed and implemented.

	\subsection{Extensions and Alternatives}\label{S4.5extensions}
	The introduced components of the three-step framework can be seen as the base approach, which may be further modified to better align the solutions with individual preferences or other requirements. In the following, we briefly present some possible extensions and alternative formulations.
	
	\subsubsection{Device Feasibility}
	While the initialization of the ideal power exchange with the market does not allow for many alternatives, the min-cost flow problem can easily be modified. In particular, the cost structure of the graph can be used to represent a number of different secondary objectives and preferences. However, it should be noted that the cost structure should still be in line with the hierarchical approach of trading, i.e. the cost of arcs connecting nodes in $\mathcal{S}$ and $\mathcal{D}$ should be lower than the cost of arcs incident to nodes in $\mathcal{F}$ and $m$. In addition, the costs of arcs incident to $m$ should be the highest among all. Given these restrictions, possible extensions could encompass:
	\begin{itemize}
		\item Use the cost of arcs between microgrids to express some mutual preference between the microgrids. This could be due to similarity or the geographical distance between neighborhoods.
		\item Use the cost of arcs to express the willingness of a microgrid to participate in peer-to-peer trades and to use its flexibility. This may be of particular interest for microgrids in $\mathcal{F}$. Microgrids that are willing to trade their flexibility to help other microgrids may associate a very small additional cost with their incident arcs, while microgrids that prefer to keep their flexibility for future time slices may impose a larger additional weight.
		\item Use the cost of arcs to connect the min-cost flow problem with the underlying electricity grid. Each trade between microgrids has an impact on the outcome of the power flow computations by increasing and decreasing the demand at two points in the electricity grid. Assuming that these two points are close by in the grid, the affected part of the grid is rather small, leading to a higher probability of the flow still being feasible. Therefore, another idea is to use the costs of the arcs connecting the microgrids with each other to represent some distance measure on the underlying electricity grid. Thereby, trades within the same branch of the grid may be preferred.
	\end{itemize}
	
	Next to the cost structure of the arcs, the graph itself may also be used to represent limitations of the peer-to-peer trading scheme. Within the currently presented graph, all microgrids are connected to each other. An interesting idea is to restrict the arcs to better incorporate the underlying grid or communication structure.

	\subsubsection{Power Flow Computation}
	Instead of using the DC power flow computation, which only approximates the physical properties of the underlying electricity grid, more accurate AC power flow computations could be used for the first computation. In recent years, new numerical solution techniques have been developed, which considerably speed up the AC power flow computation \cite{ACPFClaurent2022Giraldoetal}. Therefore, even AC power flow computations may be employed to verify if line limits are respected.
	
	\subsubsection{Parameter Updates}
	Regarding the distribution of power among the various devices of a single microgrid, the proposed solution maximizes the PV usage and tries to allocate the remaining power equally among all batteries and EVs. However, there are other ways to distribute the power among the devices:
	\begin{itemize}
		\item A first approach is to use the target values of batteries and EVs as a reference for the distribution of the remaining power, instead of using the flexibility of the devices. Based on these values, the goal of the allocation could be to minimize the maximal deviation over all devices.
		\item A second approach focuses again on the flexibility of the devices. As mentioned in Section \ref{S3Models}, the flexibility of the batteries and EVs depends on various aspects, whereby the initial energy at the beginning of the time slice is one of them. Therefore, the decision of how much power to use for the device in a time slice may have an impact on the flexibility of the device in the following time slice. Hence, we distribute the remaining power in such a way that the flexibility of the devices for the next time slice is maximized.
	\end{itemize}
	
	Finally, a very interesting decision is the update of the power exchange level $X_i$. As the main goal is to minimize fluctuations in the power exchange with the market, this value often does not change. However, using this parameter to decide how much to trade with the market at a certain time slice also allows the microgrids to participate in balancing markets. Thereby, microgrids could support the TSO with (local) imbalance regulations. Given an external signal, the update of parameter $X_i$ may be fixed to a specific value for a certain time window (a couple of seconds to minutes). Depending on the deviation from the originally desired power exchange level and the duration, the target values for the batteries and EVs may need to be adjusted to ensure a feasible solution.

	\section{Numerical Study}\label{S5Analysis}
	In this section, we test and evaluate the proposed real-time control approach and analyze and discuss the numerical results achieved for multiple underlying grid structures. We first introduce and explain the used data in Section \ref{s5.1Data}, before testing the above-presented algorithm in detail in Section \ref{s5.2Results}. We also compare the results to two other algorithms. The first one represents the case of a naive real-time control, whereby each device charges or discharges with its intended power level, and any surplus of PV is simply fed into the grid. The second algorithm represents the opposite and is an offline algorithm, that extends the system model (\ref{stssmObj})-(\ref{stssmLastConstr}) by considering all time slices at once and has access to the (aggregated) PV and household load data of each microgrid for the whole time horizon. Thereby, it serves as a lower bound on the possible objective values of the real-time control approach.\\
	
	The algorithms are implemented in Python 3.9, and Gurobi 10.0 is used to solve the mathematical optimization models on a standard laptop with an Intel Core i5-8250U CPU and 8 GB RAM.
	
	\subsection{Data}\label{s5.1Data}
	The required data to test the proposed real-time algorithm can be split up into three parts: The electricity grid data, the microgrid data for the day-ahead operation problem, and the microgrid data for the real-time algorithm. Hereby, the electricity grid data directly determines the number and the sizes of the considered microgrids. Each microgrid is defined by a set of (aggregated) PV generation and household load profiles, as well as a communal battery and EVs.

	\subsubsection{Electricity Grid Data}
	To test and analyze the presented real-time control approach, we use some of the smaller Matpower grid examples, presented in \cite{Matpower2011Zimmermanetal}. Within these grids, the number of microgrids ranges from 3 to 41, while the total aggregated load is between 0.315 and 1.25 MW per grid. Note, that in the original grid data, each grid had multiple buses producing electricity. Due to the structure of our real-time approach, we assume that only one of these buses serves as the connection to the grid. To also be able to test the real-time balancing algorithm for a larger number of microgrids and a higher load, we use the MV power distribution system as presented in \cite{NorwegianMVGrid2023Sperstadetal}. This grid has a radial structure, which fits well with the assumption of having one connection to the electricity markets. The grid consists of 124 buses, and 123 branches, whereas not every bus is associated with a load. In total, there are 54 buses with a non-zero load, which correspond to the microgrids, and the aggregated load equals 6.407 MW. Further detailed information on this grid can be found in \cite{NorwegianMVGrid2023Sperstadetal}. Table \ref{tab:Grids} provides an overview of the tested grids, including a summary of their microgrid configuration.\\
	
	\begin{table}[]
		\centering
		\begin{tabular}{ c|c|c|c|c|c|c } 
			Name & buses & branches & MGs & size MGs & households & $\sum$ load\\
			\hline
			\textit{case9} & 9 & 9 & 3 & 100 - 138 & 349 & 0.315 MW\\ 
			\textit{case14} & 14 & 20 & 11 & 3 - 104 & 283 & 0.259 MW\\ 
			\textit{case57} & 57 & 80 & 41& 1 - 418 & 1312 & 1.25 MW\\
			\textit{caseNW} & 124 & 123& 54 & 16 - 506 & 7091 & 6.407 MW\\
		\end{tabular}
		\caption{Overview of tested grids, including the MG configuration}
		\label{tab:Grids}
	\end{table}
	
	\subsubsection{Microgrid Data}
	\subsubsection*{Day-Ahead}
	As mentioned, each microgrid is associated with a specific bus in the electricity grid. In particular, the number of households within a microgrid directly relates to the load associated with the corresponding bus. We assume a peak average power consumption of 0.9 kW per household and 15-minute time slot. This results in microgrids consisting of 1 to 506 households for the various MV grids.\\
	
	The households in a microgrid can be characterized by their (inflexible) demand profiles, their EVs with corresponding arrival and departure times and demands, as well as their PV systems. The demand profile for the households is based on the average Dutch power profile for a whole year (\cite{NEDUDemandProfiles}), while the EVs are modeled after the VW ID.4 model with internal charging and discharging limits of 11 kW, charging and discharging efficiencies of 0.95 and a capacity of 58 kWh. In addition to the individual households, the microgrid also has access to a communal battery with a total capacity of 42 kWh and (internal) charging and discharging limits of 15 kW and charging and discharging efficiencies of 0.95.
	
	\subsubsection*{Real-Time}
	The mentioned power profiles of the household loads as well as the PV generation are based on 15-minute time slots and represent the energy usage or generation during a time slot. In the real-time algorithm, on the other hand, we need a finer time granularity to know how the energy usage or generation spreads within the 15-minute time slots. For the PV generation, we use the data set published in \cite{PVdataset2022Pombo}, which offers a granularity of 1 second. Furthermore, in \cite{householddataset2015Tjaden}, 74 synthetic household demand profiles for a year with 1-second granularity are presented. For a single 15-minute time slot, we normalize these profiles and then scale them with the energy demand or generation of the time slot, for which we test the real-time algorithm.

	\subsection{Results and Analysis}\label{s5.2Results}
	In the following, we present and analyze the results achieved by the real-time control algorithm in detail and compare them to the solutions of the naive control approach as well as the optimal offline model. We start with a comparison of the results of the three approaches and highlight the advantages and disadvantages of the real-time approach. We then proceed with an analysis of how the three approaches can deal with uncertainty, before focusing on the computation times of the real-time approach.
	
	\subsubsection{Comparison to Benchmark Algorithms}
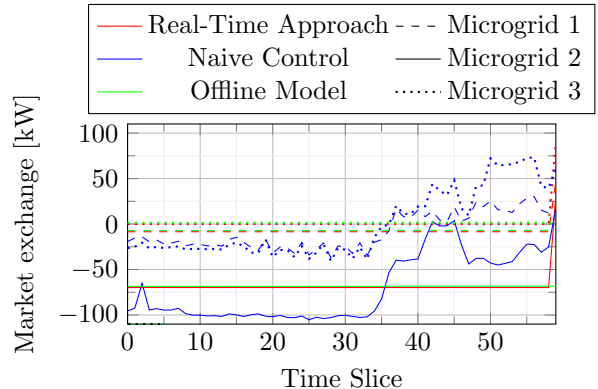
\begin{figure}
		\centering
		\pgfplotsset{every axis legend/.append style={at={(0.5, 1.6)}, anchor = north, legend columns = 2}}
		\begin{tikzpicture}
			\begin{axis}[xmin=0, xmax=59, ymin=-110, ymax=110,
				xlabel={Time Slice},
				ylabel={Market exchange [kW]},
				grid = both,
				minor tick num = 1,
				major grid style = {lightgray},
				minor grid style = {lightgray!25},
				height = 0.35\textwidth,
				width = 0.60\textwidth]
				\addplot[red]{-111};
				\addplot[black, dashed]{-111};
				\addplot[blue]{-111};
				\addplot[black]{-111};
				\addplot[green]{-111};
				\addplot[black, thick, dotted]{-110};
				\addplot[red, dashed]table[col sep = comma, x={T}, y={RTC4}]{9bus15secMarket.csv};
				\addplot[red]table[col sep = comma, x={T}, y={RTC6}]{9bus15secMarket.csv};
				\addplot[red, thick, dotted]table[col sep = comma, x={T}, y={RTC8}]{9bus15secMarket.csv};
				\addplot[blue, dashed]table[col sep = comma, x={T}, y={NAIVE4}]{9bus15secMarket.csv};
				\addplot[blue]table[col sep = comma, x={T}, y={NAIVE6}]{9bus15secMarket.csv};
				\addplot[blue, thick, dotted]table[col sep = comma, x={T}, y={NAIVE8}]{9bus15secMarket.csv};
				\addplot[green, dashed]table[col sep = comma, x={T}, y={OPT4}]{9bus15secMarket.csv};
				\addplot[green]table[col sep = comma, x={T}, y={OPT6}]{9bus15secMarket.csv};
				\addplot[green, thick, dotted]table[col sep = comma, x={T}, y={OPT8}]{9bus15secMarket.csv};
				\legend{Real-Time Approach,  Microgrid 1, Naive Control,Microgrid 2, Offline Model,Microgrid 3}
			\end{axis}
		\end{tikzpicture}
		\caption{Power exchange with the market over time for microgrids 1, 2 and 3 for the three different approaches for electricity grid \textit{case9} and $\Delta t = 15$ sec.}
		\label{FigBaseMarket}
	\end{figure}

\begin{figure}
		\centering
		\pgfplotsset{every axis legend/.append style={at={(0.5, 1.6)}, anchor = north, legend columns = 2}}
		\begin{tikzpicture}
			\begin{axis}[xmin=0, xmax=59, ymin=-7.5, ymax=7.5,
				xlabel={Time Slice},
				ylabel={Peer-to-Peer exchange [kW]},
				grid = both,
				minor tick num = 1,
				major grid style = {lightgray},
				minor grid style = {lightgray!25},
				height = 0.35\textwidth,
				width = 0.60\textwidth]
				\addplot[red]{-11};
				\addplot[black, dashed]{-11};
				\addplot[blue]{-11};
				\addplot[black]{-11};
				\addplot[green]{-11};
				\addplot[black,thick, dotted]{-11};
				\addplot[red, dashed]table[col sep = comma, x={T}, y={RTC4}]{9bus15secP2P.csv};
				\addplot[red]table[col sep = comma, x={T}, y={RTC6}]{9bus15secP2P.csv};
				\addplot[red, thick, dotted]table[col sep = comma, x={T}, y={RTC8}]{9bus15secP2P.csv};
				\addplot[blue, dashed]table[col sep = comma, x={T}, y={NAIVE4}]{9bus15secP2P.csv};
				\addplot[blue]table[col sep = comma, x={T}, y={NAIVE6}]{9bus15secP2P.csv};
				\addplot[blue, thick, dotted]table[col sep = comma, x={T}, y={NAIVE8}]{9bus15secP2P.csv};
				\addplot[green, dashed]table[col sep = comma, x={T}, y={OPT4}]{9bus15secP2P.csv};
				\addplot[green]table[col sep = comma, x={T}, y={OPT6}]{9bus15secP2P.csv};
				\addplot[thick, green, dotted]table[col sep = comma, x={T}, y={OPT8}]{9bus15secP2P.csv};
				\legend{Real-Time Approach,  Microgrid 1, Naive Control,Microgrid 2, Offline Model,Microgrid 3}
			\end{axis}
		\end{tikzpicture}
		\caption{Aggregated power exchange between microgrids over time for microgrids 1, 2 and 3 for the three different approaches for electricity grid \textit{case9} and $\Delta t = 15$ sec.}
		\label{FigBaseP2P}
	\end{figure}
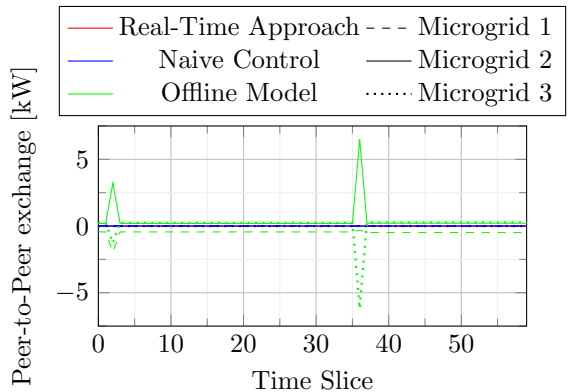

	In the following, we present some first results for the individual microgrids of MV grid \textit{case9} (see Table \ref{tab:Grids}) for the three different algorithms (see Figure \ref{FigBaseMarket}). Next to the market exchange within the different time slices (15 seconds) of the overall time horizon of 15 minutes, the aggregated peer-to-peer trading between the microgrids for each time slice is given (see figure \ref{FigBaseP2P}).\\
	
	Comparing the market exchanges of the microgrids of the naive control with the optimal offline results highlights the need for a real-time control approach to minimize power fluctuations on an MV grid level (see Figure \ref{FigBaseMarket}). The fluctuations in the power profile of the naive control approach are caused by the short-term fluctuations and mismatch between the PV generation and the household loads. The offline model with its perfect knowledge on the other hand can react to these fluctuations by a clever scheduling of charging and discharging of batteries and EVs, which leads to a perfectly flat profile. The real-time control algorithm also aims to minimize these fluctuations and it achieves this goal for all but the last time slice, as can be seen in Figure \ref{FigBaseMarket}.\\
	
	Similarly to the optimal offline model, the real-time algorithm also makes use of storage devices, such as the communal battery and the EVs. Hereby, it aims to supply power in time slices, where the PV generation and the ideal market exchange are not sufficient to cover the household demand and to store power in times when there is too much supply. However, these additional charging and discharging decisions cause a loss of energy due to the charging and discharging (in-)efficiencies, which was not accounted for in the day-ahead solution. To still ensure the demand-supply balance throughout the whole time slot, more energy than planned has to be bought for these approaches. The main difference between the real-time control and the offline model approach lies in the way the energy lost due to the inefficiency of batteries and EVs is bought. While the offline model can use its knowledge of the whole time horizon to evenly distribute the additional power among all time slices, the real-time control approach only realizes its shortcoming at the last time slice, leading to a sudden power peak, see Figure \ref{FigBaseMarket}. This also explains the slight differences between the flat power profiles of the real-time approach and the optimal offline model.\\
	
	However, a simple increase of the parameter $X_i$ to account for the energy loss may already counter this problem. It should also be noted, that the naive control approach does not suffer from this imbalance due to additional charging and discharging. However, as it simply feeds any PV surplus into the grid, it directly suffers from a highly fluctuating power exchange profile, which is also not desirable from a market point of view.\\
	
	When focusing on the peer-to-peer exchange between microgrids, we notice that in this case, only the optimal offline model makes use of the option (see Figure \ref{FigBaseP2P}). In general, the real-time control approach only makes use of the peer-to-peer trading option when the device flexibility is not sufficient for some microgrids. The offline model, on the other hand, prefers to use peer-to-peer trading, as this minimizes the additional usage of storage devices, and thereby the unaccounted losses.

	\subsubsection{Impact of Uncertainty in PV and Household Forecasts}
	
	\begin{figure}
		\centering
		\begin{subfigure}[b]{0.45\textwidth}
			\centering
			\pgfplotsset{every axis legend/.append style={at={(2.1, 1.55)}, anchor = north east, legend columns = 2}}
			\begin{tikzpicture}
				\begin{axis}[xmin=1, xmax=60, ymin=-220, ymax=150,
					ylabel={Market exchange [kW]},
					grid = both,
					minor tick num = 1,
					major grid style = {lightgray},
					minor grid style = {lightgray!25},
					width=\textwidth]
					\addplot[red]{-230};
					\addplot[black,]{-230};
					\addplot[blue]{-230};
					\addplot[black, dashed]{-230};
					\addplot[green]{-230};
					\addplot[red, dashed]table[col sep = comma, x={T}, y={0_1.0_RTC}]{9bus15secAggregated.csv};
					\addplot[red]table[col sep = comma, x={T}, y={1_1.0_RTC}]{9bus15secAggregated.csv};
					\addplot[blue, dashed]table[col sep = comma, x={T}, y={0_1.0_NAIVE}]{9bus15secAggregated.csv};
					\addplot[blue]table[col sep = comma, x={T}, y={1_1.0_NAIVE}]{9bus15secAggregated.csv};
					\addplot[green, dashed]table[col sep = comma, x={T}, y={0_1.0_OPT}]{9bus15secAggregated.csv};
					\addplot[green]table[col sep = comma, x={T}, y={1_1.0_OPT}]{9bus15secAggregated.csv};
					\legend{Real-Time Approach,  Scenario 1, Naive Control, Scenario 2, Offline Model}
				\end{axis}
			\end{tikzpicture}
			\caption{PV realization factor 1.0.}
			\label{FigUncertainty1.0}
		\end{subfigure}
		\hfill
		\begin{subfigure}[b]{0.45\textwidth}
			\centering
			\pgfplotsset{every axis legend/.append style={at={(0.99, 1.265)}, anchor = north east, legend columns = 2}}
			\begin{tikzpicture}
				\begin{axis}[xmin=1, xmax=60, ymin=-220, ymax=150,
					grid = both,
					minor tick num = 1,
					major grid style = {lightgray},
					minor grid style = {lightgray!25},
					width=\textwidth]
					\addplot[red]{-230};
					\addplot[black, dashed]{-230};
					\addplot[blue]{-230};
					\addplot[black]{-230};
					\addplot[green]{-230};
					\addplot[red, dashed]table[col sep = comma, x={T}, y={0_1.03_RTC}]{9bus15secAggregated.csv};
					\addplot[red]table[col sep = comma, x={T}, y={1_1.03_RTC}]{9bus15secAggregated.csv};
					\addplot[blue, dashed]table[col sep = comma, x={T}, y={0_1.03_NAIVE}]{9bus15secAggregated.csv};
					\addplot[blue]table[col sep = comma, x={T}, y={1_1.03_NAIVE}]{9bus15secAggregated.csv};
					\addplot[green, dashed]table[col sep = comma, x={T}, y={0_1.03_OPT}]{9bus15secAggregated.csv};
					\addplot[green]table[col sep = comma, x={T}, y={1_1.03_OPT}]{9bus15secAggregated.csv};
				\end{axis}
			\end{tikzpicture}
			\caption{PV realization factor 1.03.}
			\label{FigUncertainty1.03}
		\end{subfigure}
		\begin{subfigure}[b]{0.45\textwidth}
			\centering
			\pgfplotsset{every axis legend/.append style={at={(0.99, 1.265)}, anchor = north east, legend columns = 2}}
			\begin{tikzpicture}
				\begin{axis}[xmin=1, xmax=60, ymin=-220, ymax=150,
					xlabel={Time Slice},
					ylabel={Market exchange [kW]},
					grid = both,
					minor tick num = 1,
					major grid style = {lightgray},
					minor grid style = {lightgray!25},
					width=\textwidth]
					\addplot[red]{-230};
					\addplot[black, dashed]{-230};
					\addplot[blue]{-230};
					\addplot[black]{-230};
					\addplot[green]{-230};
					\addplot[red, dashed]table[col sep = comma, x={T}, y={0_1.1_RTC}]{9bus15secAggregated.csv};
					\addplot[red]table[col sep = comma, x={T}, y={1_1.1_RTC}]{9bus15secAggregated.csv};
					\addplot[blue, dashed]table[col sep = comma, x={T}, y={0_1.1_NAIVE}]{9bus15secAggregated.csv};
					\addplot[blue]table[col sep = comma, x={T}, y={1_1.1_NAIVE}]{9bus15secAggregated.csv};
					\addplot[green, dashed]table[col sep = comma, x={T}, y={0_1.1_OPT}]{9bus15secAggregated.csv};
					\addplot[green]table[col sep = comma, x={T}, y={1_1.1_OPT}]{9bus15secAggregated.csv};
				\end{axis}
			\end{tikzpicture}
			\caption{PV realization factor 1.1.}
			\label{FigUncertainty1.1}
		\end{subfigure}
		\hfill
		\begin{subfigure}[b]{0.45\textwidth}
			\centering
			\pgfplotsset{every axis legend/.append style={at={(0.99, 1.265)}, anchor = north east, legend columns = 2}}
			\begin{tikzpicture}
				\begin{axis}[xmin=1, xmax=60, ymin=-220, ymax=150,
					xlabel={Time Slice},
					grid = both,
					minor tick num = 1,
					major grid style = {lightgray},
					minor grid style = {lightgray!25},
					width=\textwidth]
					\addplot[red]{-230};
					\addplot[black, dashed]{-230};
					\addplot[blue]{-230};
					\addplot[black]{-230};
					\addplot[green]{-230};
					\addplot[red, dashed]table[col sep = comma, x={T}, y={0_1.4_RTC}]{9bus15secAggregated.csv};
					\addplot[red]table[col sep = comma, x={T}, y={1_1.4_RTC}]{9bus15secAggregated.csv};
					\addplot[blue, dashed]table[col sep = comma, x={T}, y={0_1.4_NAIVE}]{9bus15secAggregated.csv};
					\addplot[blue]table[col sep = comma, x={T}, y={1_1.4_NAIVE}]{9bus15secAggregated.csv};
					\addplot[green, dashed]table[col sep = comma, x={T}, y={0_1.4_OPT}]{9bus15secAggregated.csv};
					\addplot[green]table[col sep = comma, x={T}, y={1_1.4_OPT}]{9bus15secAggregated.csv};
				\end{axis}
			\end{tikzpicture}
			\caption{PV realization factor 1.4.}
			\label{FigUncertainty1.4}
		\end{subfigure}
		\caption{Aggregated power exchange with the market over time for the three different approaches for electricity grid \textit{case9} and $\Delta t = 15$ sec and various levels of PV realization.}
		\label{FigUncertainty}
	\end{figure}
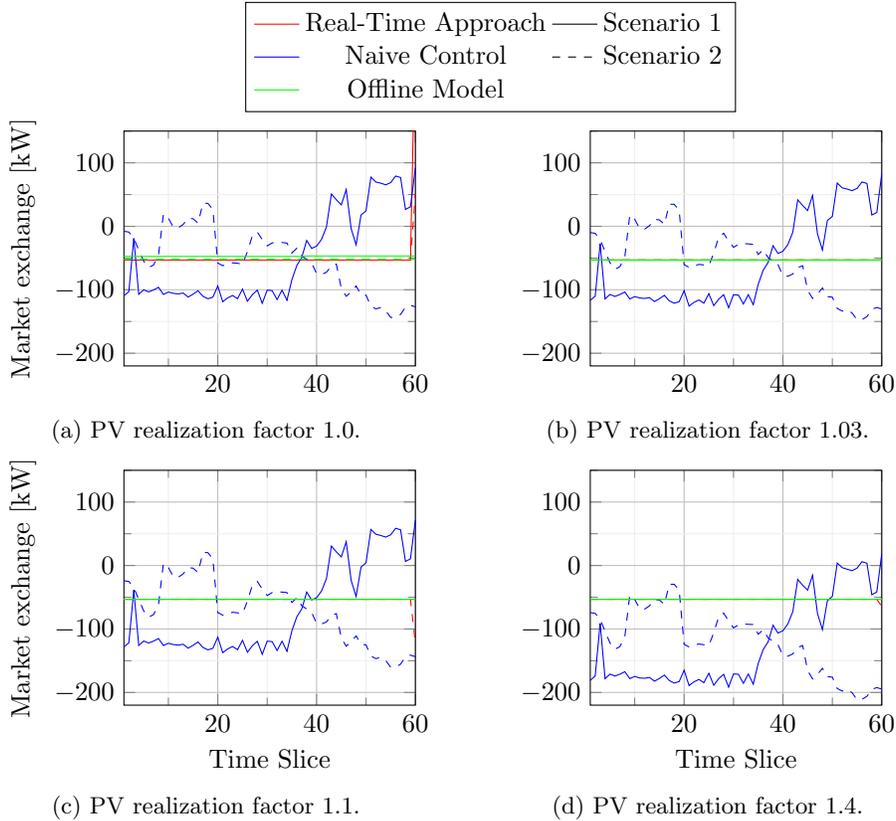
	
	A similar problem to the unaccounted loss of energy may occur when considering the effect of uncertainty on the algorithms. In the following, we analyze the impact of uncertainty on the real-time control algorithm in detail. We differentiate between two types of uncertainty: The day-ahead uncertainty, which is represented by the uncertain total amount of energy needed by households or generated by PV systems for a specific time slot, and the real-time uncertainty, which reflects how the household loads or PV generation are spread within a time slot.\\
	
	Starting with the day-ahead uncertainty, we focus mainly on the uncertain PV generation. In a small area, as represented by our considered MV grids, there is a high correlation between the realizations of PV generation. In contrast, we assume that the household load uncertainty is independent for each household. Therefore, even for the smallest grid with about 300 households (see Table \ref{tab:Grids}), we see that the individual uncertainty cancels out to a high degree. Hence, the prediction of the aggregated household load within a grid is quite accurate and we do not focus on this uncertainty during the analysis.\\
	
	In Figure \ref{FigUncertainty1.0}, it can be seen that the problem of the unaccounted loss of energy appears even if the day-ahead prediction of the PV uncertainty is perfect. However, the problem gets even worse, when there is an overestimation of the PV generation in the day-ahead operation phase. In that case, demand and supply are already out of balance over the whole time slot even without considering the unaccounted losses due to the additional usage of storage devices. The main problem with the current real-time control algorithm is that it only realizes this overestimation of the PV generation just before the last time slice. It then suddenly has to react to this information, leading to a large peak in consumption, which may even be grid-infeasible (see Figure \ref{FigUncertainty1.0}). An underestimation of the PV generation, on the other hand, can easily be dealt with. Even a small underestimation of the PV generation leads to a situation, in which the real-time control approach can use this additional generation to balance out the unaccounted energy loss (see Figure \ref{FigUncertainty}), and thereby once again obtain an optimal solution. This also holds true for large underestimations, as can be seen in Figure \ref{FigUncertainty1.4}.\\
	
	Summarizing, the results of the real-time approach strongly depend on the quality of the PV forecasts in the day-ahead stage, however this is mainly one-sided. The algorithm encounters difficulties in finding a feasible solution when faced with an overestimation of the PV generation over the whole time slot, however, it can easily deal with an underestimation of the PV generation. This observation emphasizes the need to account for uncertainty in the day-ahead operation problem. A suitable candidate is (adaptive) robust optimization with its focus on feasibility independent of the uncertainty realization, which thereby often results in an underestimation of the actual PV generation.\\
	
	The second type of uncertainty, namely the real-time uncertainty, is already taken care of by the design of the real-time control algorithm. Due to the online nature of the algorithm, it makes decisions based only on the information from the past or the current time slice. Hence, it is independent of any forecast of future generation or demand. In Figure \ref{FigUncertainty}, the results for two different real-time uncertainty realizations are given. Scenario 1 corresponds to a realization, in which the PV generation has a peak in the beginning, while the aggregated household load has its peak towards the end of the time slot. This is a scenario, in which the oversupply of power in the early time slices can be used to charge batteries and EVs and use that stored energy at later time slices to cover for the higher demand. Scenario 2 is the opposite, it starts with a peak in demand and ends with a peak in PV generation. Theoretically, this should be a more complex setting for any approach, however, if the initial state of charge of EVs and batteries is sufficient, the approach is able to cover the early demand peak. Hence, the real-time approach is able to deal with real-time uncertainty to a large extent.

	\subsubsection{Running Time and Scalability}
	
	Given the real-time nature of the proposed algorithm, its running time is of high importance to ensure that decisions can be made in time. To test and analyze the running time of the algorithm, we first theoretically analyze the running time of each individual component, presented in Section \ref{S4RTCA}. The second part is composed of a numerical simulation, which supports the theoretical conclusions and shows the potential of the algorithm for a real-world implementation.\\
	
\begin{figure}
		\centering
		\begin{tikzpicture}
			\begin{axis}
				[boxplot/draw direction=y,
				ylabel={running time per iteration (sec)},
				xtick={1,2,3,4},
				xticklabels={\textit{case9}, \textit{case14},\textit{case57},\textit{caseNW}},
				ymajorgrids,
				yminorgrids,
				xmajorgrids,
				ytick = {0, 0.05, 0.1, 0.15},
				yticklabels = {0.00, 0.05, 0.10, 0.15},
				minor tick num = 1,
				major grid style = {lightgray},
				minor grid style = {lightgray!25},
				height = 0.35\textwidth,
				width = 0.60\textwidth]
				\addplot+[boxplot prepared={
					median=0.0060,
					upper quartile=0.0062,
					lower quartile=0.0052,
					upper whisker=0.012,
					lower whisker=0.0030
				},red] coordinates {};
				\addplot+[boxplot prepared={
					median=0.00501,
					upper quartile=0.006,
					lower quartile=0.00499,
					upper whisker=0.0190,
					lower whisker=0.0040
				},blue] coordinates {};
				\addplot+[boxplot prepared={
					median=0.017,
					upper quartile=0.0197,
					lower quartile=0.016,
					upper whisker=0.109,
					lower whisker=0.010
				},green] coordinates {};
				\addplot+[boxplot prepared={
					median=0.034,
					upper quartile=0.037,
					lower quartile=0.031,
					upper whisker=0.142,
					lower whisker=0.027
				},orange] coordinates {};
			\end{axis}
		\end{tikzpicture}
		\caption{Boxplot statistics on the running time per time slice for various grid types for a time slice length of 1 second.}
		\label{FigRTBoxPlot}
	\end{figure}
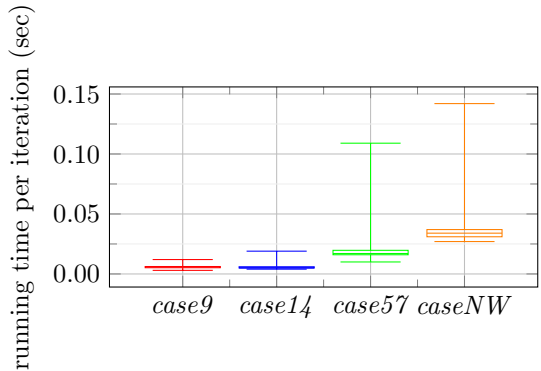

	Following Algorithm \ref{algRTC}, we have up to five computation steps per time slice:
	\begin{itemize}
		\item Computation of device bounds: Computing the device bounds only requires a few arithmetic calculations, which can be done efficiently. In addition, these calculations can be made in parallel for each microgrid and device, leading to a very fast computation. Note, that for the results presented in Table \ref{tab:RTGrids} and Figures \ref{FigRTAverage} and \ref{FigRTBoxPlot} only the computations for the microgrids are run in parallel, while the computations for the devices within each microgrid are still run in sequence. 
		\item Min-cost flow problem: The min-cost flow problem only depends on the number of microgrids and is independent of the number of households and thereby the size of the microgrids. Even in large MV grids, the number of microgrids is usually relatively small. Therefore, this problem can be solved efficiently, using either an LP solver or well-known strongly polynomial algorithms (see e.g., \cite{MCF1985Tardos}, \cite{MCF1988Orlin}).    
		\item Calculation of power flow: Given the focus on DC power flow equations, the calculation can be reduced to a simple matrix-vector multiplication. The size of the matrix (respectively the vector) depends on the number of lines and buses in the electricity grid. It is possible to further speed up the computation by ignoring the buses, which are not connected to microgrids.
		\item Repair algorithm: In its current form, the repair algorithm is modeled as an LP with a quadratic objective function. Given suitably chosen weights in the objective function, it is well known that such problems can be solved in weakly polynomial time, \cite{QPcomplexity1980Kozlovetal}. Once again, the size of the resulting mathematical model only depends on the number of microgrids as well as the number of lines within the grid and thereby should be rather small compared to the number of devices or households.
		\item Parameter update: The parameter update, as presented in Section \ref{S4.4PU}, can be done in parallel for each microgrid. Within each microgrid, the cave-filling problem can be solved in linear time on the number of devices \cite{CFPeff2016Naiduetal}. The computation of the parameter $X_i$ is a simple numerical calculation following equation (\ref{eqPU4}). 
	\end{itemize}
	
	\begin{table}[]
		\centering
		\begin{tabular}{ c|c|c|c|c|c|c|c|c } 
			& \multicolumn{2}{c|}{\textit{case9}} & \multicolumn{2}{c|}{\textit{case14}} & \multicolumn{2}{c|}{\textit{case57}} & \multicolumn{2}{c}{\textit{caseNW}}\\
			$\Delta t$ & RTC & OA & RTC & OA & RTC & OA & RTC & OA \\
			\hline
			60 & 0.082 & 1.122 & 0.077 & 1.925 & 0.239 & 71.180 & 0.483 & 525.601\\
			30 & 0.157 & 2.457 & 0.142 & 4.091 & 0.473 & 144.532 & 1.068 & 1099.56\\
			15 & 0.330 & 6.528 & 0.309 & 9.037 & 0.968 & 298.101 & 2.002 & 2158.19\\
			10 & 0.498 & 11.020 & 0.437 & 15.153 & 1.458 & 473.66 & 3.372 & 3492.01\\
			5 & 0.996 & 38.708 & 0.925 & 46.152 & 3.036 & - & 6.395 & -\\
			1 & 5.310 & - & 5.046 & - & 16.723 & - & 33.012 & -\\
		\end{tabular}
		\caption{Total running time (in sec.) for the real-time control (RTC) approach and the offline approach (OA) for various time slice lengths (in sec.) and electricity grids. Entries with - did not terminate due to a shortage of memory.}
		\label{tab:RTGrids}
	\end{table}
	
	\begin{figure}
		\centering
		\pgfplotsset{every axis legend/.append style={at={(0.9, 1.41)}, anchor = north east, legend columns = 2}}
		\begin{tikzpicture}
			\begin{axis}[xmin=0.5, xmax=6.5, ymin=0, ymax=0.045,
				xlabel={Time Slice Length (sec)},
				ylabel={avg. running time per iteration (sec)},
				xtick = {1,2,3,4,5,6},
				xticklabels={60,30,15,10,5,1},
				grid = both,
				minor tick num = 1,
				major grid style = {lightgray},
				minor grid style = {lightgray!25},
				legend cell align={left},
				height = 0.35\textwidth,
				width = 0.60\textwidth]
				\addplot[red, mark = triangle*]{-1};
				\addplot[blue, mark = square*]{-1};
				\addplot[green, mark = otimes*]{-1};
				\addplot[orange, mark = x]{-1};
				\addplot[red, mark=triangle*] table {
					1  0.0055
					2  0.0052
					3  0.0055
					4  0.0055
					5  0.0055
					6  0.0059
				};
				\addplot[blue, mark=square*] table {
					1  0.0051
					2  0.0047
					3  0.0052
					4  0.0049
					5  0.0051
					6  0.0056
				};
				\addplot[green, mark=otimes*] table {
					1  0.0160
					2  0.0158
					3  0.0161
					4  0.0162
					5  0.0168
					6  0.0185
				};
				\addplot[orange, mark=x] table {
					1  0.032
					2  0.036
					3  0.033
					4  0.037
					5  0.036
					6  0.037
				};
				\legend{\textit{case9}, \textit{case14}, \textit{case57}, \textit{caseNW}}
			\end{axis}
		\end{tikzpicture}
		\caption{Average running time per iteration (in sec.) of the real-time control approach for the four different electricity grids.}
		\label{FigRTAverage}
	\end{figure}
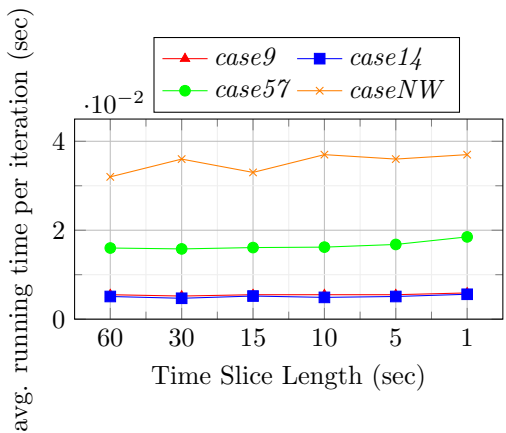
	
	Table \ref{tab:RTGrids} and Figures \ref{FigRTAverage} and \ref{FigRTBoxPlot} show the running times of the real-time control approach for various time slice lengths and grids. Hereby, Table \ref{tab:RTGrids} compares the total running times of the real-time control algorithm and the offline model approach, while Figure \ref{FigRTAverage} focuses on the average running time per time slice of only the real-time control algorithm. Figure \ref{FigRTBoxPlot} shows for a time slice length $\Delta t$ of one second some additional statistics on the running time per time slice beyond the average value.\\
	
	Comparing the real-time control algorithm with the offline model, we notice that the real-time control algorithm scales much better with the number of time slices, respectively the time slice length (see Table \ref{tab:RTGrids}). Figure \ref{FigRTAverage} supports this conclusion in that the running time per time slice is nearly constant for all grids. The running time of the offline model on the other hand does not follow such a clear linear trend for smaller time slices and instances with a very large number of time slices often cannot be solved due to a shortage of memory. In addition, even instances that are solvable need 14 to 1000 times more computation time compared to the real-time control approach.\\
	
	In general, the running time of the real-time control algorithm per time slice mainly depends on the number of microgrids and the corresponding number of households per microgrid, but is nearly independent of the time slice length (see Figure \ref{FigRTAverage}). This insight supports the theoretical analysis of the individual components, in which the running time of the components depends on either the number of microgrids, the number of devices per microgrid, or the size of the underlying electricity grid. However, the running time of an individual time slice may also be affected by which components are used. As shown in Algorithm \ref{algRTC}, not every component is always used. Both, the min-cost flow problem as well as the repair algorithm are only run if either the device flexibility of microgrids is not sufficient or the resulting solution violates branch limits. These cases mostly do not show up, leading to very low computation times in the vast majority of iterations (see Figure \ref{FigRTBoxPlot}), although outliers may appear. Nevertheless, even these outliers are small enough for the algorithm to be employed in a real-time setting. It should be noted that the overhead due to communication has not been considered and must be added to the running times in case of a real-world implementation.

	\section{Discussion and Conclusion}\label{S6Conclusion}
	This work aims to design a grid-aware real-time control algorithm, which uses the decisions made at the day-ahead market to guide its real-time actions. We created a three-step framework, which allows a set of microgrids to jointly control their balancing actions to maintain a flat power exchange profile with the electricity market. The framework is built up of multiple separate components, which can be set up according to individual preferences or computational requirements. Due to an aggregation step at the microgrid level, privacy-relevant information and data are not shared outside the microgrid, which may increase the acceptance for real-world employment. A case study on multiple MV grids shows promising results w.r.t. the running time and the objective of the algorithm. A comparison to an optimal offline model revealed that given some light assumptions, the real-time control approach obtained an optimal solution in most cases.\\
	
	The comparison to the naive control strategy, in which all devices simply act according to their planned power levels, also highlights the importance of having a real-time control approach to implement the day-ahead solution in a predictable manner. In particular, it is not desirable from a market perspective to not have any control within the 15-minute time slots. Due to the increased share of non-renewable energy sources as well as the increased load due to EVs or heat pumps, the mismatch between demand and supply leads to highly fluctuating power profiles, which may also synchronize within a region. This situation calls for a control approach, which can deal with the imbalance in demand and supply in real-time. Our proposed algorithm solves this imbalance on three different layers. In the first layer, each microgrid uses the flexibility of its own devices to ensure a balance. If this local mechanism is not sufficient, the second layer enables microgrids to trade with each other. Thereby, neighboring microgrids can help each other with their remaining flexibility on a regional level. In the third level, microgrids can increase or decrease their power exchange with the market to temporarily deal with the remaining demand or supply. This hierarchical structure increases the self-consumption of renewable energy generation within microgrids and promotes the trading between neighboring microgrids. The advantages of these incentives are the reduced usage of the electricity grid and the connection of producers and consumers on a local level.\\
	
	The analysis also revealed another interesting aspect in the interplay between day-ahead operation and real-time control approaches, which is often neglected or ignored. Implementing the planned (average) power exchange level with the market throughout the whole time slot often requires the usage of storage devices such as batteries beyond the planned day-ahead usage. Assuming imperfect batteries, losses due to the additional charging and discharging, may disrupt the planned demand-supply balance. There are two levels, where this issue can be addressed. In the real-time layer, the additional demand (to cover the losses) may be realized by increasing the target power exchange level slightly. However, this may directly lead to an imbalance in the markets, in case every microgrid requires more power than planned. The second option is to account for the losses due to real-time balancing already in the day-ahead operation problem. This avoids the problem of increasing the actual load of every microgrid during the real-time control and decreases the deviation in the demand-supply balance.\\ 
	
	Based on the achieved results and interesting insights, various interesting research directions arise. In a first step, a more detailed study on the interplay between day-ahead operation and real-time control approaches is necessary to explore and evaluate the different options to deal with the additional battery usage and its accompanying energy losses. In Section \ref{S4.5extensions}, a short overview of possible extensions and alternatives to the proposed components is provided. A further study of their performance in the three-step framework could produce interesting and valuable insights beyond our analysis. A last research direction may focus on the implemented components of the framework. Due to the real-time aspect of this algorithm, the running time of each component is of high importance, and tailor-made algorithms may further speed up the overall computation time.

	\section*{Acknowledgements}
	This research is supported by the Netherlands Organization for Scientific Research (NWO) Grant 645.002.001.
	
	\printbibliography
	
\end{document}